\documentclass[reprint,superscriptaddress,amsmath,amssymb,prb]{revtex4-1}

\usepackage{graphicx}
\usepackage{dcolumn}
\usepackage{bm}
\usepackage{color}

\begin{document}

\preprint{APS/123-QED}

\title{Multiple superconducting phases driven by pressure in the topological insulator GeSb$_4$Te$_7$}

\author{W. Zhou}
\email{wei.zhou@cslg.edu.cn}
\affiliation{School of Electronic and Information Engineering, Changshu Institute of Technology, Changshu 215500, China}

\author{B. Li}

\affiliation{Information Physics Research Center, Nanjing University of Posts and Telecommunications, Nanjing 210023, China}

\author{Y. Shen}
\affiliation{Key Laboratory of Low-Dimensional Quantum Structures and Quantum Control of Ministry of Education, Department of Physics
and Synergetic Innovation Center for Quantum Effects and Applications, Hunan Normal University, Changsha 410081, China}

\author{J. J. Feng}
\affiliation{Center for High Pressure Science and Technology Advanced Research, Shanghai 201203, China}

\author{C. Q. Xu}
\affiliation{School of Physical Science and Technology, Ningbo University, Ningbo 315211, China}

\author{H. T. Guo}
\affiliation{School of Electronic and Information Engineering, Changshu Institute of Technology, Changshu 215500, China}

\author{Z. He}
\affiliation{School of Electronic and Information Engineering, Changshu Institute of Technology, Changshu 215500, China}

\author{B. Qian}
\email{njqb@cslg.edu.cn}
\affiliation{School of Electronic and Information Engineering, Changshu Institute of Technology, Changshu 215500, China}

\author{Ziming Zhu}
\affiliation{Key Laboratory of Low-Dimensional Quantum Structures and Quantum Control of Ministry of Education, Department of Physics
and Synergetic Innovation Center for Quantum Effects and Applications, Hunan Normal University, Changsha 410081, China}

\author{Xiaofeng Xu}
\email{xuxiaofeng@zjut.edu.cn}
\affiliation{Department of Applied Physics, Zhejiang University of Technology, Hangzhou 310023, China}

\date{\today}

\begin{abstract}
Tuning superconductivity in topological materials by means of chemical substitution, electrostatic gating or pressure is thought to be an effective route towards realizing topological superconductivity with their inherent Majorana fermions, the manipulation of which may form the basis for future topological quantum computing. It has recently been established that the pseudo-binary chalcogenides ($ACh$)$_m$($Pn_2$$Ch_3$)$_n$ ($A$ = Ge, Mn, Pb, etc.; $Pn$ = Sb or Bi; $Ch$= Te, Se) may host novel topological quantum states such as the quantum anomalous Hall effect and topological axion states. Here we map out the phase diagram of one member in this series, the topological insulator candidate GeSb$_4$Te$_7$ up to pressures of $\sim$35 GPa, through a combination of electrical resistance measurements, Raman spectroscopy as well as first-principles calculations. Three distinct superconducting phases emerge under the pressure above $\sim$11 GPa, $\sim$17 GPa and $\sim$31 GPa, which are accompanied by concomitant structural transitions, evidenced from the changes in the Raman modes. The first-principles calculations validate the existence of a topological insulating state at ambient pressure and predict two possible structural transitions at 10 GPa and 17 GPa, in agreement with the experimental observations. Overall, our results establish the GeSb$_4$Te$_7$ family of materials as a fertile arena for further exploring various topological phenomena, including topological phase transitions and putative topological superconductivity.
\end{abstract}

\maketitle

\section{Introduction}
Since the first theoretical proposals of two-dimensional (2D) and three-dimensional (3D) topological insulators (TIs), the material realizations of nontrivial electronic band topology with unconventional surface states and peculiar electromagnetic responses have become one of the most fascinating fields in solid-state physics \cite{Hasan, Qi, Yan1, Ando}. Subsequently, time-reversal-symmetric, nonsymmorphic topological insulators; mirror Chern insulators; Dirac, Weyl and nodal-line semimetals; as well as higher-fold degenerate fermions and high-order TIs have all been put forward theoretically and been verified experimentally \cite{Yan2, Bernevig, Tang, Xu, Po,Fuliang}. The success achieved in the past decade has been further expedited by recently introduced theories of topological quantum chemistry and symmetry-based indicators (SIs) that have in particular facilitated high-throughput searches for topological phases of matter \cite{Xu,Po,Vergniory,Bradlyn}. Although over 50\% of all known materials exhibit symmetry-indicated nontrivial topology \cite{Bernevig22}, only very limited proportion of them become superconductors at low temperatures, either at ambient pressure or under hydrostatic pressure. This hinders the search for intrinsic topological superconductors in stoichiometric compounds, thereby presenting challenges in exploiting Majorana fermions for topological quantum computing.

In a large number of currently known topological materials, the layered binary chalcogenides $Pn_2Ch_3$ ($Pn$= Bi, Sb; $Ch$= Te, Se) were among the first validated TIs, both theoretically and experimentally \cite{Hasan,Ando}. These Bi$_2$Te$_3$-type compounds have tetradymite-like layered structures comprised of Te-Bi-Te-Bi-Te quintuple layers (5L,QLs) that are stacked by van der Waals forces along the $c$-axis (Fig. 1a) \cite{Eremeev}. This weak van der Waals force facilitates the chemical doping or intercalation. For example, the MnTe bilayer can readily intercalate the QLs, forming the Te-Bi-Te-Mn-Te-Bi-Te septuple layers (7L,SLs) (Fig. 1b). With these QLs and SLs as the building blocks, new van der Waals compounds ($ACh$)$_m$($Pn_2Ch_3$)$_n$ ($A$ = Ge, Mn, Pb, etc.; $Pn$ = Sb or Bi; $Ch$= Te, Se) can be synthesized \cite{Li,Deng,Otrokov,Hu,Chowdhury,Da Silva,Park}. Taking MnBi$_2$Te$_4$ ($m$=1, $n$=1) as an example, the structure is constructed by the stacking of SLs (Fig. 1b), whereas in MnBi$_4$Te$_7$ ($m$=1, $n$=2), it consists of alternating stacking of QLs and SLs along the $c$-axis, forming naturally grown superlattices (Fig. 1c). Interestingly, most members in this family were found to be 3D TIs \cite{Ando,Otrokov,Hu,Muff,Huan}, thereby providing a tunable platform to observe unusual topological phenomena.

\begin{figure*}
\includegraphics[width=17cm]{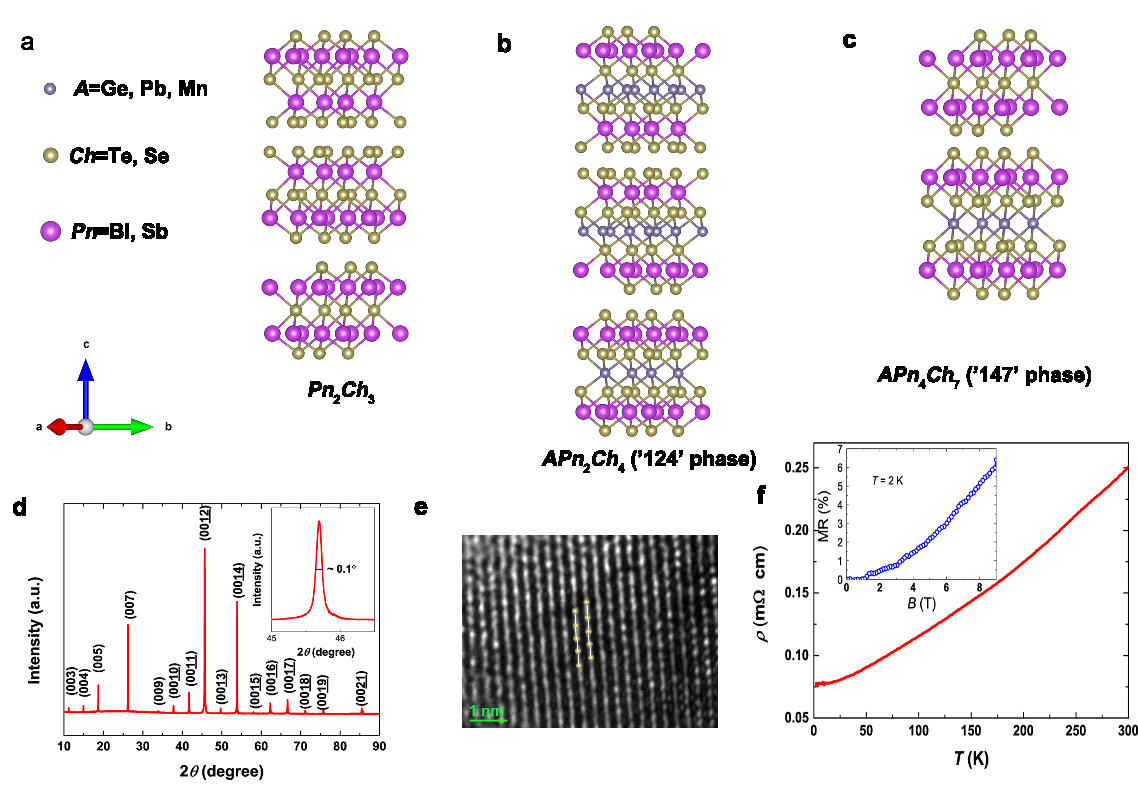}
\caption{\label{fig1} \textbf{a}, \textbf{b}, and \textbf{c}, Crystal structures of $Pn_2Ch_3$, $APn_2Ch_4$ ('124' phase), and $APn_4Ch_7$ ('147' phase) at ambient pressure. The structures are drawn using the software VESTA \cite{Momma}. \textbf{d}, The ambient XRD pattern of a GeSb$_4$Te$_7$ single crystal. Inset is an enlarged view of the (00\underline{12}) peak with a FWHM of 0.1$^\circ$. \textbf{e}, TEM image of the crystal. \textbf{f}, Temperature dependence of the resistivity at ambient pressure. Inset shows the magnetoresistance (MR) measured at \textit{T} = 2 K at ambient pressure.}
\end{figure*}

Previously, much effort has been devoted to realizing superconductivity in the ($ACh$)$_m$($Pn_2Ch_3$)$_n$ compounds, and pressure has been demonstrated to be an important tool for tuning structural phase transitions and superconductivity thereof \cite{Kong, Zhu, Matsumoto, Song, Greenberg, Hen}. For example, in Sb$_2$Te$_3$, four superconducting phases were revealed under pressures up to 32 GPa that are correlated with four distinct structures \cite{Zhu}. In SnSb$_2$Te$_4$, superconductivity was found to be gradually enhanced with pressure up to 33 GPa \cite{Song}. In both crystalline GeSb$_2$Te$_4$ and amorphous GeSb$_2$Te$_4$, two superconducting phases were observed for $P <$ 40 GPa and the maximum superconducting transition temperature was observed at 20-30 GPa \cite{Greenberg, Hen}.

Notably, the magnetic members in family such as MnBi$_2$Te$_4$ and MnBi$_4$Te$_7$, have attracted more interest recently, with the desire to observe the quantum anomalous Hall (QAH) effect and axion insulator state with quantized topological magnetoelectric effect \cite{Li,Deng,Otrokov,Hu,Lee,Li1,Li2,Vidal}. Following the first observation of the QAH effect in Cr-doped (Bi,Sb)$_2$Te$_3$ TI thin films \cite{Chang}, the focus has shifted towards intrinsic magnetic topological materials that are thought to provide a cleaner platform for studying emergent magnetic topological states. Indeed, QAH was observed in MnBi$_2$Te$_4$ thin flakes with an odd number of SLs \cite{Deng}. On the other hand, pressure has also been employed to tune the ground states of these magnetic topological materials. Specifically, it was reported that pressure induces several structural transitions in MnBi$_2$Te$_4$ and MnBi$_4$Te$_7$, which change their topological properties simultaneously \cite{Pei}. However, no superconductivity was observed in both systems up to $\sim$50 GPa \cite{Pei}. More recently, pressure-induced superconductivity was observed in the Mn-based '147' antiferromagnet MnSb$_4$Te$_7$ with a maximum $T_c$ of 2.2 K at 50 GPa \cite{Pei2}. A natural question arises as to the possible role played by the magnetic fluctuations in its superconductivity since the element Mn is generally magnetic. In particular, whether the magnetic fluctuations are energetically favorable or adverse for the formation of superconductivity in this topological family is unclear.

With all these in mind, we performed high pressure study of the nonmagnetic counterpart GeSb$_4$Te$_7$ that is isostructural to MnSb$_4$Te$_7$. We reveal three distinct superconducting phases under pressure above 11 GPa, 17 GPa and 31 GPa, which are concomitant with three structural transitions, different from the pressure phase diagrams reported in Sb$_2$Te$_3$, SnSb$_2$Te$_4$, GeSb$_2$Te$_4$, etc\cite{Kong, Zhu, Matsumoto, Song, Greenberg, Hen}. Importantly, the superconducting transition temperature $T_c$ in pressurized GeSb$_4$Te$_7$ is a factor of 4 higher than that in its pressurized magnetic homologue MnSb$_4$Te$_7$. This significantly enhanced $T_c$ suggests that the magnetic fluctuations may have adverse effects on the Cooper pair formation in this '147' phase of pseudo-binary systems, although both GeSb$_4$Te$_7$ and MnSb$_4$Te$_7$ show unambiguous topological features.

\begin{figure*}
\includegraphics[width=17cm]{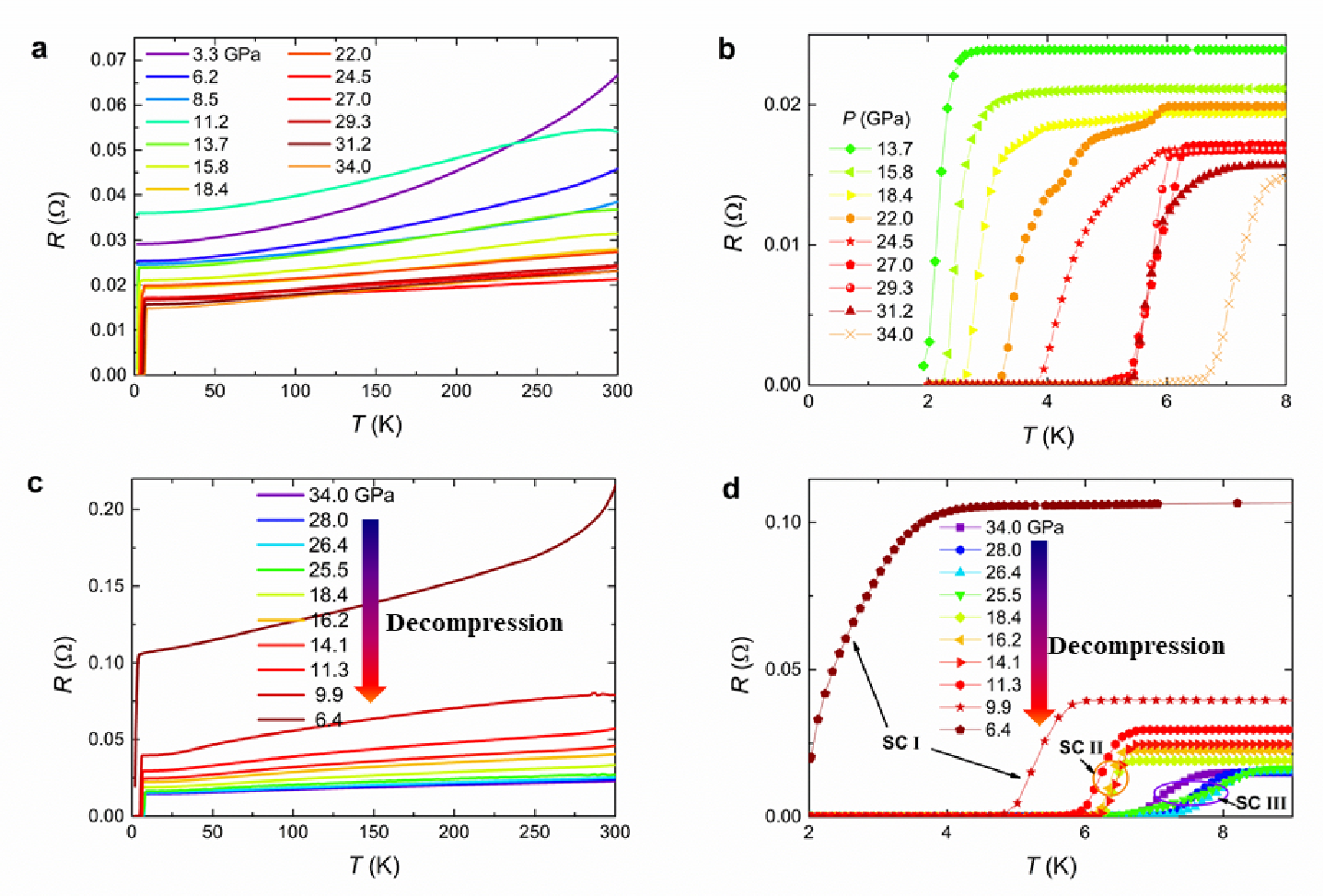}
\caption{\label{fig1} \textbf{a}, Temperature-dependence of the resistance $R(T)$ for different pressures measured during the compression  process. \textbf{b}, an enlarged view of \textbf{a} in the low-temperature region. \textbf{c}, $R(T)$ curves for different pressures measured during the decompression process. \textbf{d}, a low temperature enlargement of panel \textbf{c}.}
\end{figure*}

\section{Experiment}

Single crystals of GeSb$_4$Te$_7$ were grown by the flux method reported in Ref. \cite{Rohr}. Single crystal and powder x-ray diffraction (XRD) measurements were performed at room temperature using a diffractometer with Cu $K\alpha$ radiation. The actual composition of the single crystals was characterized by an energy dispersive x-ray spectroscopy (EDX) analyzer equipped on a scanning electronic microscope (SEM) and a transmission electron microscope (TEM). The EDX results confirm that the atomic ratio for Ge: Sb: Te is very close to 1: 4: 7. The specific heat data were measured on a Physical Property Measurement System (PPMS-9, Quantum Design). The transport data under pressure were acquired from measurements in a diamond anvil cell (DAC) with NaCl as the pressure transmitting medium. High-pressure Raman spectroscopy data were collected by a Renishaw InVia Raman system with a laser wavelength of 532 nm.

The electronic band structure calculations were performed using the full-potential linearized augmented plane wave (FP-LAPW) method implemented in the WIEN2K code \cite{Blaha}. We employed a tight-binding dependent package WannierTools for topological properties investigations \cite{Wu2}. The iterative Green's function was used for obtaining the surface state spectrum. We used the in-house developed crystal structure prediction package CRYSTREE to search for the stable structures of GeSb$_4$Te$_7$ under pressure \cite{Wu}. We then employed the Quantum ESPRESSO (QE) package \cite{Giannozzi} with Perdew-Burke-Ernzerhof (PBE) exchange-correlation functional to compute the enthalpies of the predicted structures under various pressures. The cutoff values for charge density and the wave function were set to be 600 Ry and 60 Ry, respectively.

\section{Results}

\begin{figure}
\includegraphics[width=8.6cm]{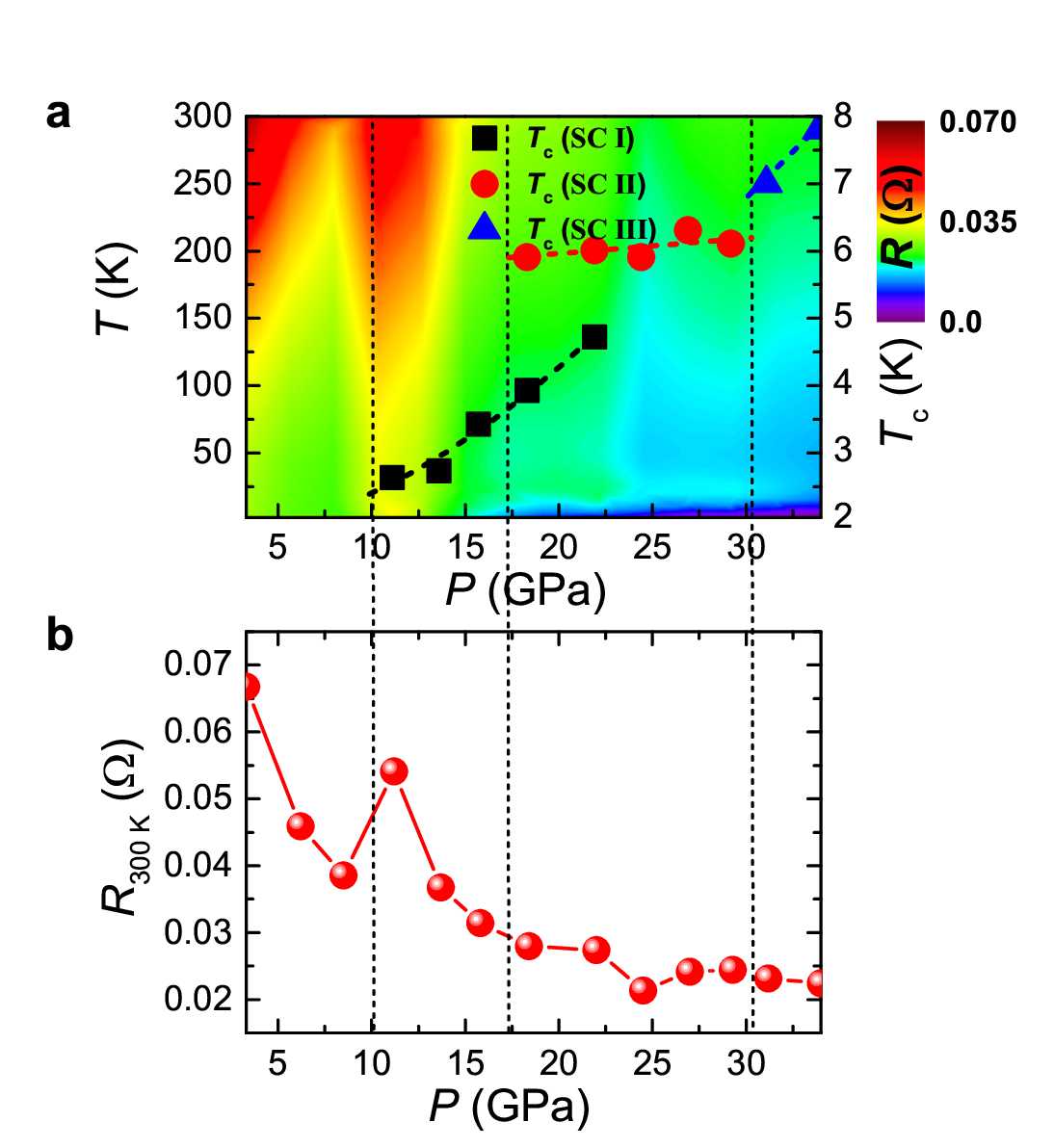}
\caption{\label{fig3} \textbf{a}, Pressure-dependent phase diagram based on the data measured during the compression process. The right vertical coordinate denotes the onset superconducting transition temperature $T_c$. \textbf{b}, Pressure dependence of the resistance at 300 K.}
\end{figure}

\begin{figure*}
\includegraphics[width=17cm]{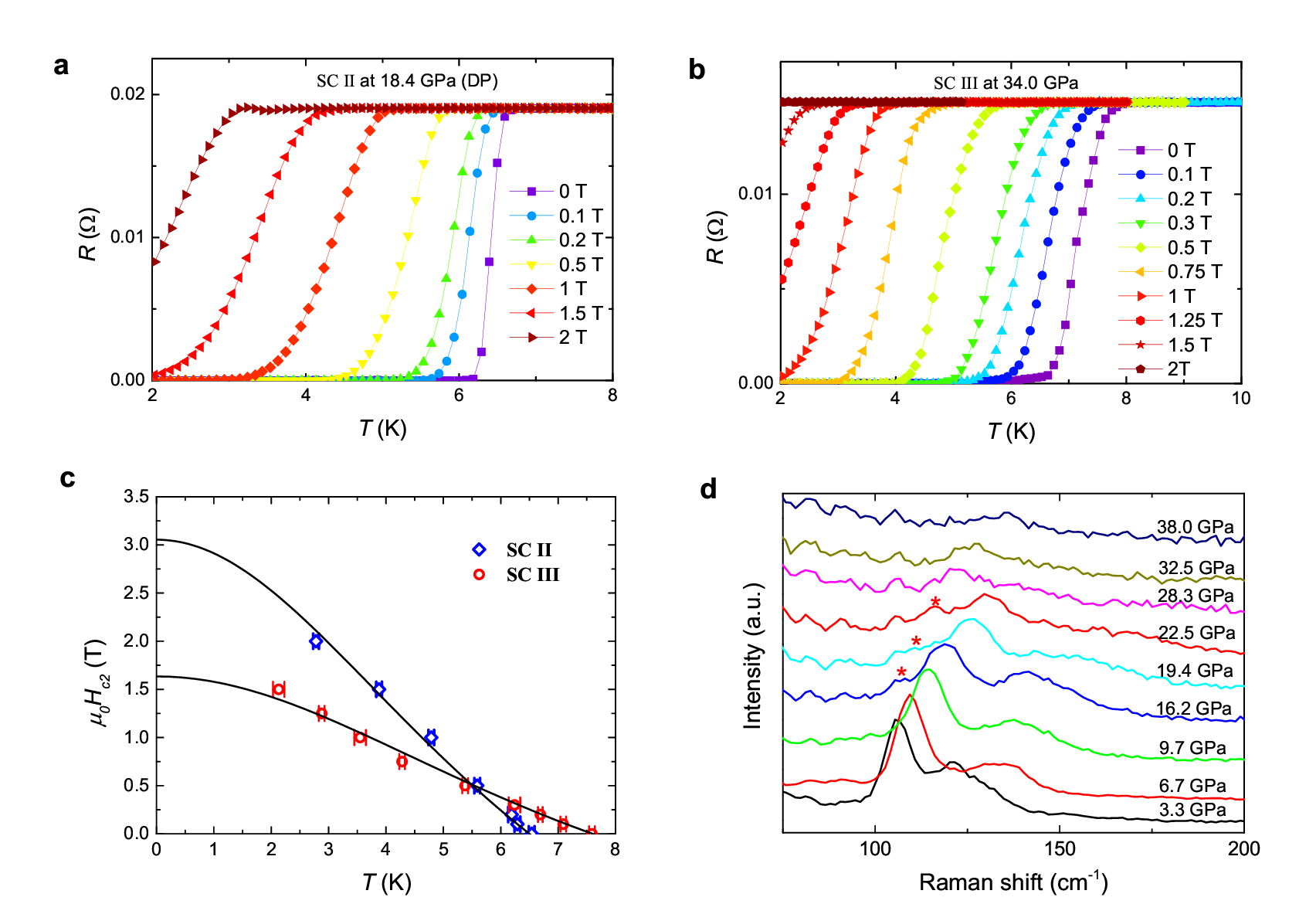}
\caption{\label{fig4} \textbf{a} and \textbf{b}, The superconducting transitions measured under different magnetic fields. \textbf{c}, Temperature dependence of the upper critical field $\mu_0H_{c2}$. The solid lines are fits based on the Ginzburg-Landau formula, that is, $\mu_0H_{c2}$($T$)=$\mu_0H_{c2}$(0)(1-$t^2$)/(1+$t^2$ ), where $t=T/T_c$, yielding $\mu_0H_{c2}$(0)= 3.1 T for SC II and $\mu_0H_{c2}$(0)= 1.6 T for SC III. \textbf{d}, Raman spectra measured under different pressures.}
\end{figure*}

Figure 1d shows the single crystal XRD pattern for one GeSb$_4$Te$_7$ crystal under ambient conditions. Only the reflections from the (00$l$) planes are observed, indicating that the crystallographic $c$-axis is normal to the sample surface. The sharp peaks demonstrate the high quality of the single crystal, with a FWHM of the (00\underline{12}) peak of about 0.1$^{\circ}$, exemplified in the inset of Fig. 1d. As shown in Fig. S1 of the Supplementary Information (SI), all peaks of the powder XRD from the pulverized crystals can be well indexed by the trigonal $P\bar{3}m1$ (No. 164) space group \cite{Supplementary}. The calculated lattice parameters are $a = 4.23$ {\AA} and $c = 23.82$ {\AA}, in excellent agreement with the published results \cite{Petrov}. The TEM image further demonstrates the high quality of the single crystal, as illustrated in Fig. 1e. The temperature dependent resistivity shows typical metallic behavior from room temperature all the way down to 2 K (Fig. 1f), with no superconductivity observed in this temperature range. The residual resistivity ratio ($R_{300 K}/R_{2 K}$) is rather small ($\sim$3.3), indicating significant electron scattering. The inset of Fig. 1f shows the magnetoresistance (MR) measured at 2 K. In comparison with many TIs, the MR value is relatively small, which may also be related to significant electron scattering. The heat capacity characterization of the sample shows no evident anomaly below 200 K (Fig. S2 of Ref. \cite{Supplementary}). The fit based on the electron and phonon contributions $C(T) = \gamma_{n}T+\beta T^{3}+\eta T^{5}$ yields a Sommerfeld coefficient $\gamma_{n}$ = 2.43 mJ/mol K$^{2}$, and a Debye temperature $\Theta_{D}$ = 187.5 K. This small Sommerfeld coefficient implies weak electron correlations in this material.

\begin{figure*}
\includegraphics[width=17cm]{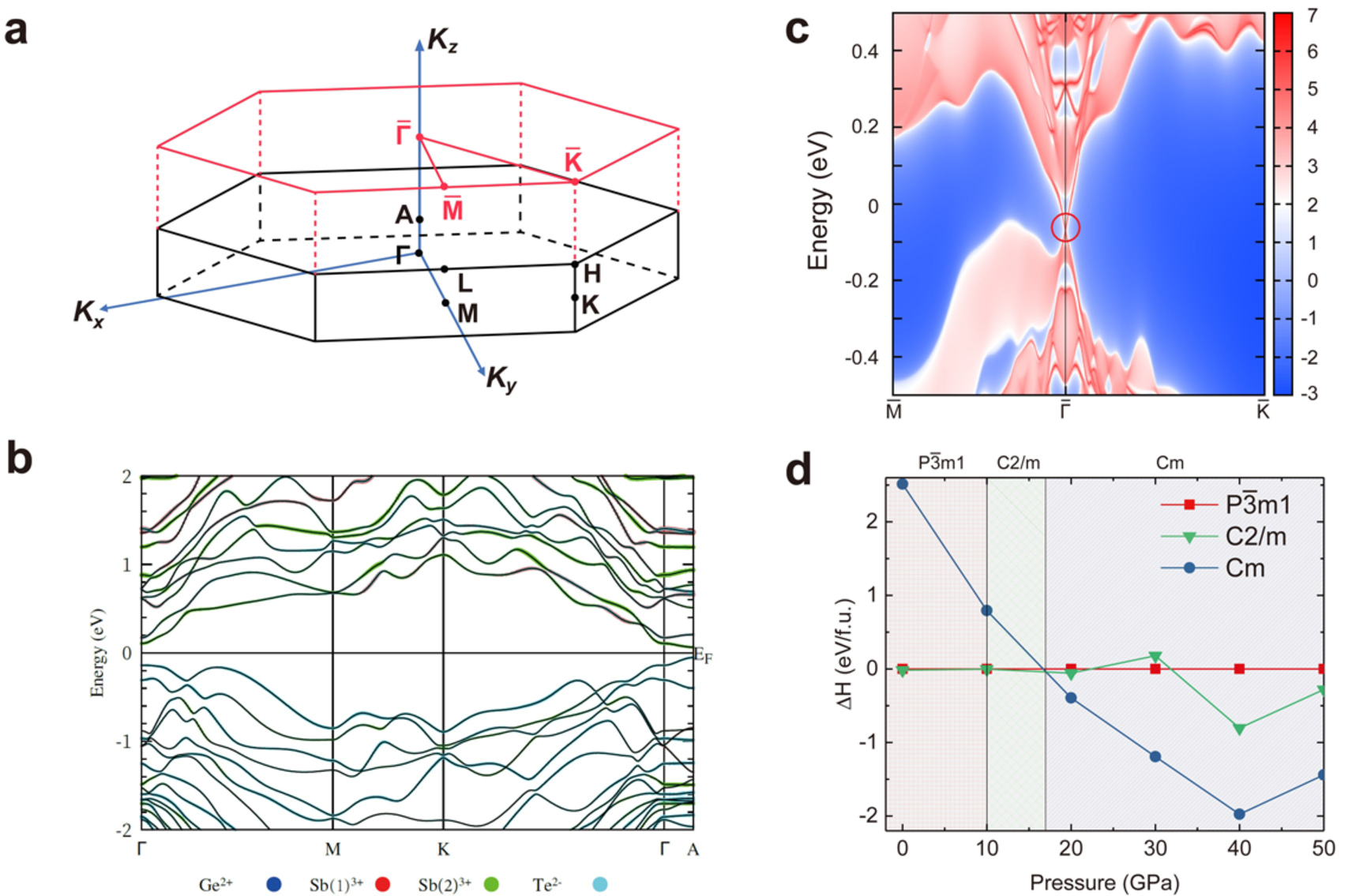}
\caption{\label{fig5} \textbf{a,} Bulk and (001)-surface Brillouin zone of GeSb$_4$Te$_7$ for the $P\bar{3}m1$ phase. \textbf{b,} The electronic band structure of the $P\bar{3}m1$ phase colored by elemental characters at ambient pressure. \textbf{c,} Surface state for the (001)-surface along high-symmetry directions. \textbf{d,} Calculated enthalpies of different phases as a function of pressure. The enthalpy of the $P\bar{3}m1$ phase at the corresponding pressure is taken as the reference enthalpy, i.e., $\Delta H=H(phase)-H(P\bar{3}m1)$.}
\end{figure*}

The $R(T)$ profiles measured under various pressures upon compression are shown in Figure 2. With increasing pressure, the $R(T)$ curves are gradually suppressed up to 8.5 GPa, above which the $R(T)$ curve shows a sudden increase (see 11.2 GPa data). Simultaneously, in the low-$T$ region, the $R(T)$ curve ($P$ = 11.2 GPa) shows a small resistance drop, indicating incipient superconductivity. As $P$ further increases, the superconducting transition becomes prominent and zero resistance can be observed. Meanwhile, the normal state $R(T)$ curve shifts downwards again. As seen in Fig. 2b, the superconducting transition gradually moves toward higher temperatures as $P$ changes from 13.7 GPa to ~24.5 GPa. However, for $P > $24.5 GPa, the superconducting transition barely changes with increasing $P$. Above 31.0 GPa, a large enhancement of $T_c$ can be seen, reaching a maximum value of 8 K at 34 GPa, the highest pressure measured in this study. In parallel, Figs. 2c and 2d show the resistance during the pressure decompression process, measured with the same electrical contacts as in the pressure compression course. With decreasing pressure, the above-mentioned main features in the $R(T)$ curves and $P$-dependent superconducting transitions can be overall reproduced, with the exception that superconductivity can now be seen at much lower pressures. Specifically, at the lowest pressure of $\sim$6.4 GPa measured upon decompression, the superconducting transition can still be observed below $\sim$3.8 K, while when compressing, the critical pressure for observing superconductivity is approximately $\sim$11.2 GPa.

The pressure dependences of the $R(T)$ curves, the superconducting transition temperatures as well as the room-temperature resistances during compression are summarized in the phase diagram in Figure 3. As noted, the emergence of superconductivity around 11 GPa is accompanied by a sudden increase of the resistance at 300 K, implying a possible phase transition around $\sim$11 GPa. With increasing pressure, $T_c$ is gradually enhanced at a rate of $\sim$0.21 K/GPa while the resistance at room temperature is smoothly suppressed. We designate this superconducting phase as SC I. As the pressure increases to 18.4 GPa, the resistance displays a small drop around 6 K initially, followed by a major decrease at 4 K. This two-step drop of resistance suggests two possible superconducting transitions and was also seen under 22 GPa. For $P >$ 24.5 GPa, only one SC transition can be observed and $T_c$ barely changes with pressure, displaying a $T_c(P)$ plateau with a $T_c$ value of $\sim$6 K. A similar weak pressure dependence of $T_c$ has also been observed in the pressure phase diagrams of Bi$_2$Se$_3$ and Sb$_2$Te$_3$, which was shown to be closely related to the charge carrier density \cite{Kong, Zhu}. We denote this phase as SC II. When $P >$ 31 GPa, $T_c$ shows a large increase, and $R_{300 K}$ starts to decrease again. These changes suggest a possible phase transition taking place around 31 GPa and we therefore label the superconducting phase for $P >$ 31 GPa as SC III. The room temperature resistance as a function of pressure is displayed in Fig. 3b. For the phase diagram for the decompression process, see Fig. S3 in Ref.\cite{Supplementary}.

The distinction between the SC II and SC III phases is not only manifested in their different pressure dependence of $T_c$, but also reflected in their temperature-dependent upper critical field ($\mu_0H_{c2}$) behaviors as displayed in Fig. 4c. The $\mu_0H_{c2}$ values are extracted from the superconducting transitions under different fields shown in Figs. 4a and 4b, based on the commonly-used 90\% criterion where the resistance drops to 90\% of the normal state value. It is worth noting that choosing a different criterion (e.g., 50\% criterion or zero-resistance) does not qualitatively change the overall behavior. In Fig. 4c, the fittings based on the Ginzburg-Landau formula $\mu_0H_{c2}$($T$)=$\mu_0H_{c2}$(0)(1-$t^2$)/(1+$t^2$ ), where $t=T/T_c$, lead to $\mu_0H_{c2}$(0)= 3.1 T for 18.4 GPa (SC II) and $\mu_0H_{c2}$(0)= 1.6 T for 34.0 GPa (SC III). Generally, for the same superconducting phase, a higher superconducting transition temperature often results in a higher $\mu_0H_{c2}$(0). However, here the superconductivity at 34 GPa (SC III) has a higher $T_c$ value but with a lower $\mu_0H_{c2}$(0), in contrast to the $\sim$6 K superconductivity of SC II, and therefore one can conclude that these two superconductivity phases are different in nature.

Pressure-dependent Raman spectra were measured and are displayed in Fig. 4d. At $P \sim$3.3 GPa, two distinct Raman modes at 106 cm$^{-1}$ ($E_g$) and 121 cm$^{-1}$ ($A_{1g}$) are observed \cite{Mal}. As $P$ increases, the Raman peaks move towards higher wave numbers. When $P$ is increased from 9.7 GPa to 16.2 GPa, a new vibration mode labeled by a red star emerges, suggestive of a structural phase transition. This structural phase transition is also consistent with the transition revealed from the transport measurements discussed above. When $P$ is increased from 22.5 GPa to 28.3 GPa, the main $E_g$ mode disappears, implying the occurrence of another structural phase transition, which agrees with the transition from the SC I to SC II. For $P >$ 30 GPa, the Raman peaks become very weak, which probably results from the pressure-induced amorphization or another structural transition.

The electronic band structure of the pristine phase ($P\bar{3}m1$) at ambient pressure is shown in Figure 5b. A small energy gap of $\sim$0.07 eV opens up around the Fermi level, suggesting a semiconducting ground state. The metallic behaviors revealed in the resistivity measurements indicate that the as-grown samples are naturally doped due to small off-stoichiometry or native defects. As seen, the valence bands are predominantly contributed by Te atoms whereas the conduction bands are dominated by Sb orbitals. We also studied the topological properties of the $P\bar{3}m1$ phase under the tight-binding framework. The $Z_2$ topological indices ($\nu_0$, $\nu_1$$\nu_2$$\nu_3$) can be estimated through the calculation of Wannier charge centers (WCCs) in six time-reversal invariant planes (i.e., $k_x$ = 0, $\pi$, $k_y$ = 0, $\pi$ and $k_z$ = 0, $\pi$ planes). The resultant topological index for the $P\bar{3}m1$ phase is (1, 001), which indicates that the $P\bar{3}m1$ phase is a strong topological insulator. Figure 5c illustrated the surface state spectrum of the (0 0 1) surface of GeSb$_4$Te$_7$, with evident Dirac surface states crossing the Fermi level in the bulk band gap, indicating its topological insulator properties.

To illuminate the nature of the transitions revealed in the transport and Raman spectroscopy measurements, we made an attempt to predict the crystal structure under pressure using the first-principles calculations, as shown in Fig. 5d. Here we calculated the enthalpy of different phases with increasing pressure and the most stable phase has the lowest enthalpy in general. Our results show that GeSb$_4$Te$_7$ maintains the ambient structure up to $P$ $\sim$10 GPa, above which the $C2/m$ phase becomes the most stable. This phase is energetically favorable until a pressure of $\sim$17 GPa where GeSb$_4$Te$_7$ undergoes a structural transition to the $Cm$ phase. This structural prediction may be overall consistent with the first two structural transitions seen in the experiments. However, our calculations can not resolve the third transition around 30 GPa, which merits further studies in the future. The band structures of the high-pressure $C2/m$ and $Cm$ phases are also calculated (see Fig. S4 in Ref. \cite{Supplementary}). The non-trivial topological characteristics are manifested in these pressure-induced new phases, evidenced from non-zero $Z_2$ invariants. Specifically, the calculation of the $C2/m$ phase at 10 GPa exhibits a $Z_2$ index of (1, 010), demonstrating that it is a strong topological insulator, whereas for the $Cm$ phase at 20 GPa, the calculated $Z_2$ index is (0, 111), indicating it is a weak topological material.

\section{Discussion and Conclusion}

Having established the phase diagram of the title compound under pressure, it is important to understand its topological properties and superconductivity by comparing to those in the same family of $(ACh)_m(Pn_2Ch_3)_n$. Topological surface states have been unambiguously demonstrated in spin-resolved angle-resolved photoelectron spectroscopy (ARPES) measurements on Ge(Bi$_{1-x}$Sb$_x$)$_4$Te$_7$ with $x$ ranging from $x$=0 to $x$=0.25 \cite{Muff}. With increasing Sb content, the Dirac point on the surface moves towards the Fermi level, changing from n- to p-type carriers at $x$=0.15. Albeit with no direct ARPES measurement on the end member GeSb$_4$Te$_7$ thus far, it is conceivable that it also harbors Dirac cones on its surface, as suggested from our first-principles calculations above. Recently, versatile topological phases have been reported in the '147' magnetic MnSb$_4$Te$_7$ whose topological properties can be either tuned by carrier doping or by magnetic field, through the change of magnetic configurations via the latter \cite{Huan}. On the other hand, extensive efforts have been made to tune the superconductivity in this class of materials by pressure but most of them are unsuccessful. For example, applying pressure on MnBi$_2$Te$_4$ and MnBi$_4$Te$_7$ only induces several structural transitions and the concomitant changes of topological properties, with no superconductivity being observed up to 50 GPa \cite{Pei}. In magnetic MnSb$_4$Te$_7$ with weaker magnetic interactions, evidenced from its lower magnetic transition temperatures compared with MnBi$_2$Te$_4$ and MnBi$_4$Te$_7$, superconductivity was observed when $P$ $>$ 30 GPa, reaching a maximum $T_c$ of 2 K at 50 GPa \cite{Pei2}. This does seem to suggest that magnetic fluctuations play an adverse role in the superconductivity, at least in this class of materials. This argument gets further support from the observations of superconductivity in the nonmagnetic counterpart GeSb$_4$Te$_7$ uncovered in this study, as well as the reported pressure-induced superconductivity in GeSb$_2$Te$_4$ ($T_c$=6 K at $P$=20 GPa) \cite{Greenberg,Hen}, which are both nonmagnetic.

From the structural point of view,  the '147' phase can be viewed as the combination of $Pn_2Ch_3$  and the '124' phase (see Fig. 1). Naively, given the weak van der Waals force between the quintuple layer and the septuple layer, one may expect that the pressure effect for the '147' phase would be a combination of that for the $Pn_2Ch_3$ phase and the '124' phase. However, the pressure effect on GeSb$_4$Te$_7$, as revealed in this study, is quite different from both that of Sb$_2$Te$_3$ and the '124' GeSb$_2$Te$_4$ \cite{Zhu,Greenberg,Hen}. In Sb$_2$Te$_3$, for example, superconductivity was observed under pressures as low as 4.0 GPa, and four superconducting phases have been identified up to 32 GPa \cite{Zhu}. In GeSb$_2$Te$_4$ (here we only consider the crystalline GeSb$_2$Te$_4$ phase \cite{Greenberg}), superconductivity was induced at $\sim$10 GPa and two superconducting phases were suggested with a maximum $T_c$ ($\sim$8 K) around 20-30 GPa \cite{Greenberg}. For GeSb$_4$Te$_7$, the lowest pressure for observing superconductivity is also around 10 GPa, which is comparable to that for GeSb$_2$Te$_4$ but much higher than that of Sb$_2$Te$_3$. In the pressure range of 10-30 GPa, the phase diagrams for GeSb$_2$Te$_4$ and GeSb$_4$Te$_7$ are quite similar. However, a definite third superconducting phase (SC III) is observed in GeSb$_4$Te$_7$ for $P > $30 GPa, which appears to be absent in GeSb$_2$Te$_4$.

In summary, we have performed a systematic high-pressure study of the topological insulator GeSb$_4$Te$_7$ which reveals multiple phase transitions upon compression. Three superconducting phases were uncovered in our high-pressure study up to 35 GPa, where the corresponding changes in their Raman spectroscopy suggest the presence structural transitions under pressure. However, first-principle calculations only reproduced the first two transitions and therefore point to the need for future studies to reveal the nature of all these structural transitions. This work not only provides further impetus for studying novel topological phenomena in these pseudo-binary chalcogenides $(ACh)_m(Pn_2Ch_3)_n$, but also calls for future investigations into the interplay between the topological states and the emergent superconductivity in the pressurized phases of this class of materials.

\begin{acknowledgments}
The authors thank Michael Smidman, Xiangang Wan, Dong Qian for helpful discussions. This work was sponsored by the National Natural Science Foundation of China (Grant no. 11704047, U1832147, U1932217). XX acknowledges the financial support from National Natural Science Foundation of China (Grants No.12274369 and No. 11974061).

W. Zhou and B. Li contributed equally to this work.

\end{acknowledgments}

\appendix

\end{document}